\documentclass[aps,prb,twocolumn,epsfig,floats]{revtex4}
\usepackage{graphicx,epsf,natbib}
\def\be{\begin{equation}}
\def\ee{\end{equation}}
\def\bea{\begin{eqnarray}}
\def\eea{\end{eqnarray}}

\begin{document}
%\documentstyle[aps,epsf,preprint]{revtex}

%\everymath{\rm}
%\everydisplay{\rm}
%\renewcommand{\baselinestretch}{1.5}
%\setlength{\textwidth}{6.25in}

\title{Quantum-Criticality in Dissipative Quantum Two-Dimensional XY\\
 and Ashkin-Teller  Models: Application to  the Cuprates} 
%alternative \title{Theory of a Marginal Fermi-liquid}
%alternative \title{Quantum Critical Fluctuations in Cuprates}
\author{Vivek Aji and C. M. Varma}
\affiliation{Physics Department, University of California,
Riverside, CA 92507}
\begin{abstract}
In a recent paper\cite{aji1} we have shown that the dissipation driven quantum phase transition of the 2D xy model represents a
universality class where the correlations at criticality is local in space and power law in time. Here we provide a detailed analysis of the model. The local criticality is brought about by the
decoupling of infrared singularities in space and time. The former leads to a Kosterlitz Thouless transition whereby the excitations of the transverse component of the velocity field (vortices)
unbind in space. The latter on the other hand leads to a transition among excitations (warps) in the longitudinal component of the velocity field, which
unbind in time. The quantum Ashkin-Teller model, with which the observed loop order in the Cuprates is described maps in the critical regime to the quantum xy model. We also discuss other models which are expected to have similar properties.
\end{abstract}
\maketitle

The dissipative quantum 2D xy model was  introduced \cite{SC, MPAF} to describe experiments on ultrathin granular Superconducting films, where it was observed that above a normal-state sheet resistance of order
$R_{Q} = h/4e^{2}$ the resistivity does not decrease towards $0$ even at the lowest temperatures studied. \cite{ORR}. The granular superconductor is represented by an array of islands with finite superfluid density with Josephson coupling
among the nearest neighbors. Dissipation of the Caldeira-Leggett \cite{CL} form is included, whereby the phase difference between the nearest neighbors is coupled to an external
bosonic bath. the external bath effectively act as a resistor, $R$, shunting nearest neighbor islands and energy is dissipated at the rate $V^{2}/R$, where $V$ is the voltage
drop which is proportional to the rate of change of the phase difference between the grains. A quantum phase transition occurs as the value of the resistance is tuned. For small values of the
resistance, phase slips, whereby a voltage develops across the junction between two grains, cost too much energy and so lead to an ordered
phase. On the other hand, for large value of the resistance, phase slip events are energetically favored and global superconductivity is destroyed.

In the absence of dissipation, the quantum model at $T\to 0$ maps to a 3d xy model. It was suggested that the physics in the presence of dissipation may not be continuously connected to the 3d xy universality class \cite{SC1}.
It was also suggested that there is a possibility that the singularity introduced by dissipation may be linked to a transition driven by the unbinding in time of  phase slips \cite{MPAF1}. The calculations
are done in the weak coupling regime where one considers the Josephson coupling perturbatively. The same approach was previously used to study the phase transition and correlation functions
\cite{SC,MPAF1,WAG}. Arguments were given also that the critical fluctuations due to the unbinding of the phase slips in time which have long-time singularities may be local in space \cite{SC}. However later work by some of the same authors presents a rather different picture \cite{Tewari}. Moreover none of these investigations into the dissipative transition are applicable near the critical point.

In a previous paper we have shown that the the quantum phase transition is obtained due to the proliferation of a new kind of excitation, which we termed warps, whose physics is indeed related to the phase slips in time; more specifically they represent events in time where a
change in the longitudinal component of the velocity field\cite{aji1} occur. The possibility that the longitudinal component may lead to the introduction of sources and sinks which control the
critical behavior was also recently emphasized in the context of one dimensional systems \cite{PAL,GIL}. The correct identification of the relevant degrees of freedom in higher dimensions turns out
to be interesting. The introduction of warps simplifies the analysis of the problem so that the correlation function of the order parameter can be obtained near the critical point. We find spatially local correlations with an infrared divergence ($\propto 1/t$) in temporal correlations at the critical point. We also the cross-over from the quantum critical to the quantum-regime.

The motivation for this study is that the existence of such a critical point naturally explains some of the most important features of the phase diagram of the Cuprate superconductors. Polarized neutron scattering \cite{FAQ, mook, greven, bourges_pc} in three distinct family of Cuprates have confirmed previous finding in a third family of Cuprates through dichroism in Angle 
Resolved Photoemission spectra
\cite{AK}, which reported the observation of time reversal violation in the pseudogapped phase of underdoped cuprates. In particular, the experiments lend support to the idea that the ground state in the pseudogapped regime is one with ordered current loops without loss of translation symmetry. \cite{CMV, CMV2, simon}. The low energy sector of this theory contains four states that are related by time reversal and reflection along the
$x$ or $y$ axis. Alternatively, the four states can be represented by two Ising variable in each unit cell, the $\pm$ values representing currents in the $x$ and $y$ directions. A classical
statistical mechanical model for such degrees of freedom that supports a phase transition without any divergence in the specific heat \cite{BAX, GRONSLETH} is the Ashkin-Teller model \cite{AT}. Such a model is known to
undergo a Gaussian phase transition \cite{JOS}. A quantum generalization of the classical model in terms of operators that induce fluctuations within the low energy phase space of the four
states, leads to the dissipative 2D xy model studied here.
The local criticality with a $1/t$ singularity is the 
fluctuation spectrum necessary to explain the anomalous properties in a number of properties such a resistivity, nuclear relaxation rate and optical conductivity in the normal state
near optimal doping \cite{mfl} abutting the pseudogapped phase. At long wavelengths, such correlations are directly observed by Raman scattering experiments.  Recently, we have also shown that such fluctuations couple to fermions to promote pairing in the $d$-symmetry channel. \cite{SHEK}

It is worth pointing out that a number of classical statistical mechanical models such as the 6-vertex and 8-vertex models also have Gaussian criticality in part of their phase diagram, as well as phase transitions which do not manifest any anomaly in their specific heat. Such models when augmented with ohmic dissipation do belong to the
same universality class considered here. Local quantum criticality happens also to be a hallmark of the quantum phase transitions in heavy fermions \cite{SCHRODER, SI}, although it is not clear to us how the class of models solved here may be connected to the microscopic models for heavy fermions.

This paper is organized as follows: The dissipative quantum XY model and the Villain transformation are discussed in sec II. In section III, we show how dissipation changes the critical properties of the model.  We show that the action can be written in a quadratic form in terms of two orthogonal topological excitations, vortices which interact only in space and warps which interact only in time. We also discuss the difficulties in analyzing the model based on other choice of variables. In sec IV we discuss the properties of the warps. Sec V deals with the scaling equations for the vortices and warps and the phase transition which occurs through the proliferation of warps. In sec VI we  present a detailed calculation of the fluctuation spectra near the quantum critical point. The spectra has the same form as that proposed to explain the normal state {\it marginal fermi-liquid} properties of the high $T_c$ cuprates. In sec VII we present the zero temperature phase diagram. The connection to Ashkin Teller model and cuprates is presented in sec VIII. The effect of fourfold anisotropy on the critical properties is anayzed in sec IX. In sec X we show how dissipation arises in cuprates.
In sec XI we discuss other forms of dissipation and how they may lead to different forms of singularities in the fluctuation spectra.

\section{Dissipative Quantum 2DXY Model}

The classical 2d xy model consists of  $U(1)$ degrees of freedom,
represented by an angle $\theta$, living on the sites of a regular
lattice, assumed to be a square lattice here, with a nearest
neighbor interaction of the form

\begin{equation}
H = J\sum_{\left\langle ij \right\rangle}\left[1 -
\cos\left(\theta_{i} - \theta_{j}\right)\right],
\end{equation}

\noindent where $J$ is Josephson coupling. Since a continuous symmetry cannot be spontaneously broken in two dimensions \cite{MW,PCH}, this model does not support a long range ordered phase.
Nevertheless a phase transition does occur at finite temperature where the correlation function of the order parameter $e^{\imath\theta}$ changes from exponential to power law. This is the
Kosterlitz-Thouless-Berezinskii transition \cite{BER, KT}. The physics of this phase transition is better understood in terms of the topological defects of the system. To do so we follow the
standard procedure of using the Villain transform and integrating out the phase degrees of freedom. We include the algebra here so as to make the later discussion of the quantum version easier
to follow.

\noindent The Villain transformation expresses the periodic
function in terms of a periodic Gaussian: 

\begin{widetext}
\begin{equation}\label{vill}
\exp\left[-\beta J\sum_{\left\langle ij \right\rangle}\left[1 -
\cos\left(\theta_{i} - \theta_{j}\right)\right]\right] \approx
\sum_{m_{ij}}\exp\left[-\beta J\sum_{\left\langle ij
\right\rangle}\left(\theta_{i}-\theta_{j}-2\pi
m_{ij}\right)^{2}/2\right],
\end{equation}
\end{widetext}

\noindent where $m_{ij}$ are integers that live on the links of the
original square lattice. On a square lattice we choose to identify
each site as $i=x,y$ where $x$ and $y$ are integers. We can combine
the two link variables $m_{x,y;x+1,y}$ and $m_{x,y;xy+1}$ into one
two component vector $\textbf{m}_{x,y}$ that lives on the site $x,y$
of the lattice. The choice is purely for convenience and none of the
results  depend on how one chooses to organize the degrees of
freedom of the model. Instead of introducing a new field to
linearize the quadratic term via the Hubbard-Stratonovich
transformation, we follow an alternative procedure. We expand the quadratic term and transform to Fourier
space. Keeping the leading quadratic term
$\theta_{x,y}-\theta_{x+1,y}\approx$ $- a\nabla_{x}\theta_{xy}$,
where a is the lattice constant we get

\begin{eqnarray}
\left(\theta_{xy}-\theta_{x+1,y}-2\pi m_{x,y}^{x}\right)^{2}\approx
a^{2}\nabla^{2}_{x}\theta_{xy} &+& 4\pi
a\nabla_{x}\theta_{xy}m_{x,y}^{x} \nonumber \\&+& 4\pi^{2}m_{x,y}^{x
2},
\end{eqnarray}

\noindent where $m^{x}_{x,y}$ is the $x$ component of the vector
field given by the integer $m_{x,y;x+1,y}$. Fourier transforming and
integrating out the field $\theta_{\textbf{k}}$ in terms of which
the action is quadratic we get

\begin{equation}
Z \approx \sum_{\textbf{m}}\exp\left[-4\beta J\pi^{2}\int
d\textbf{k}\left[{k^{2}\textbf{m}\cdot \textbf{m}_{k}-
\left(\textbf{k}\cdot\textbf{m}_{k}\right)^{2}\over
k^{2}}\right]\right].
\end{equation}

\noindent Combining the two terms and Fourier transforming to real
space we get

\begin{equation}\label{clvrtact}
Z \approx \sum_{\textbf{m}}\exp\left[-\beta
J\sum_{r}\sum_{r'}\left(\nabla\times\textbf{m}_{r}\right)\cdot\left(\nabla\times\textbf{m}_{r'}\right)\log\left|r-r'\right|\right].
\end{equation}

\noindent Since the vector field is two dimensional the curl is a
scalar and is the vorticity of the vector field. Defining integer
charges $\nabla\times\textbf{m}_{r}=\rho_{r}$ we get the standard
action of the two dimensional coloumb gas. On the lattice this
definition is equivalent to

\begin{eqnarray}
\rho_{x,y} &=& m_{x,y+1;x+1,y+1}-m_{x,y;x+1,y}\\
\nonumber &+& m_{x+1,y;x+1,y+1}-m_{x,y;x,y+1}.
\end{eqnarray}

\noindent In this model, a phase transition occurs from a state at
high temperatures where the vortices are free to a low temperature
state where they are bound in pairs.

We now turn to the description of the Quantum version of the 2d xy
model. First we consider the model in the absence of dissipation. The
 quantum  2d xy model is written in terms of operators  $\theta$ and its conjugate, the number
operator $n$. The Hamiltonian is of the form

\begin{equation}
H= \sum_{i} {n_{i}^{2}\over {2C}} -J\sum_{\left\langle
ij\right\rangle}\cos\left(\theta_{i}-\theta_{j}\right),
\end{equation}

\noindent where $C$ is the capacitance and $[n_{i},\theta_{j}] =
i\delta_{ij}$. Since $n_{i} = i\partial/\partial\theta_{i}$, the
first term is the kinetic energy of the phase with a mass $C$. There
is competition between the kinetic energy and the potential energy
terms, the former minimized by a state where $\theta_{i}$ is disordered
stabilizing an insulating phase while the latter minimized by a
fixed value of $\theta_{i}$ stabilizing a superconductor. The
partition function can be recast in the form

\begin{widetext}
\begin{equation}\label{q2dxy}
Z  = \int D \theta_{i}\left(\tau\right) \exp\left[-\int_{0}^{\beta}
d\tau \left(\sum_{i}{C \over
{2}}\left(\partial_{\tau}\theta_{i}^{2}\right) - J\sum_{\left\langle
ij\right\rangle}\cos\left(\theta_{i}-\theta_{j}\right)\right)\right]
\end{equation}.
\end{widetext}

\noindent  At $T=0$, the (imaginary) time direction becomes infinite in extent, and the model is in the 3d xy universality class. The dualization procedure used for the classical 2d xy model can now be implemented on the 3d xy action as well. First we discretize the imaginary time direction in
units of $\Delta\tau$ and work on a three dimensional lattice. Introducing the variables ${\bf m}$ which only live on the spatial links, the action is

\begin{eqnarray}
Z &\approx &
\sum_{\textbf{m}}\exp\left[\sum_{\textbf{k},\omega}-4\pi^{2}
J\Delta\tau \textbf{m}\cdot\textbf{m}\right.\\
\nonumber &+&
\left.{4\pi^{2}J^{2}ca^{2}\Delta\tau^{2}\left({\textbf{k}\cdot\textbf{m}}\right)^{2}
\over {\left(Ca^{2}/c\Delta\tau\right)\omega^{2}+Jca^{2}\Delta\tau
k^{2}}}\right],
\end{eqnarray}

\noindent where $c = a/\Delta\tau$. Rearranging terms as before we get

\begin{equation}\label{llm}
Z = \sum_{\textbf{m}}\exp\left[\sum_{\textbf{k},\omega}-4\pi^{2}J \left({Jc\left|\textbf{k}\times \textbf{m}\right|^{2}+ (C/c)\omega^{2}\left|\textbf{m}\right|^{2}\over {Jck^{2}+(C/c)
\omega^{2}}}\right)\right],
\end{equation}

\noindent where $J\rightarrow Ja^{2}\Delta\tau$, $C\rightarrow Ca^{2}/\Delta\tau$ and $\textbf{m}\rightarrow \textbf{m}/a$. The isotropic model is recovered when the discretization in time is
such that $\Delta\tau = \sqrt{C/J}$. Combining $\nabla\times \textbf{m}$ (which is a scalar since $\textbf{m}$ is two dimensional) and $\partial\textbf{m}/\partial\tau$ into one three
dimensional field we see that eqn.\ref{llm} is the partition function for the three dimensional loop gas model which is in the same universality class as the 3d xy model\cite{DH}.  A point to note here is that a vector field coupling via a three dimensional kernel does undergo a phase transition while a scalar field does not \cite{POL}.
Alternatively, the loop gas model undergoes a 3d xy transition while the three dimensional Coloumb gas is always in a disordered phase.

Alternate
derivations \cite{LEEFISHER} map the model  directly to the 3d xy model instead of the loop gas model. We find that  the procedure described above is preferable in the presence of dissipation. 

In the dissipative 2d xy model, one adds to eqn.\ref{q2dxy} a term which generates in the equations of motion of the phase variable a term proportional to the velocity, $\dot{\theta}$. The
dissipative part of the action is of the form \cite{NAG}

\begin{widetext}
\begin{equation}\label{diss}
S_{diss} = \int_{-\infty}^{\infty}d\tau \int_{0}^{\beta}d\tau
'\sum_{\left\langle ij
\right\rangle}\alpha\left({\left(\theta_{i}-\theta_{j}\right)\left(\tau\right)-\left(\theta_{i}-\theta_{j}\right)\left(\tau
'\right)\over {\tau - \tau '}}\right)^{2},
\end{equation}
\end{widetext}

\noindent where $\alpha=R_{Q}/R$ where $R_{Q}=h/4e^{2}$. For a
single Josephson junction the voltage drop is given by
$\partial(\theta_{1}-\theta_{2})/\partial\tau$. The power dissipated
by a resistor connecting the two superconducting grains is
$(\partial(\theta_{1}-\theta_{2})/\partial\tau)^{2}/R$. Integrating
over time one obtains total energy dissipated through the resistor.
For an array of junctions the corresponding term in the action is of
the form in eqn.\ref{diss}. Such a dissipative term can also be
derived by assuming that the phase difference across every junction
is coupled to an external bath of harmonic oscillators. Integrating
out the bath the dissipative term is obtained in Fourier space as

\begin{equation}
S_{diss} = \alpha\sum_{\omega_{n}}\sum_{\left\langle ij
\right\rangle}\left|\omega_{n}\right|^{\eta}\left|\theta_{i,\omega_{n}}-\theta_{j,\omega_{n}}\right|^{2}.
\end{equation}

\noindent The exponent $\eta$ reflects the form of the spectral density of the external oscillator bath at low frequencies. For a linear spectral density $\eta=1$ and a Ohmic dissipation is
obtained \cite{CL}. It is indeed interesting to ask what other forms of dissipation is allowed. We will postpone the discussion of the effect of such dissipative terms to section X.

\section{Excitations of the Quantum 2D XY Model}

We can now carry through the program of introducing the vector field
$\textbf{m}$ for the quantum 2d xy model with Ohmic dissipation
($\eta = 1$). The integration of the $\theta$ fields can be
performed since the dissipative term is also quadratic. The
effective model is now written as

\begin{widetext}
\begin{eqnarray}\label{dissvrt}
Z  & = & \sum_{\textbf{m}}\exp\left[\sum_{\textbf{k},\omega}-4\pi^{2} J \textbf{m}\cdot\textbf{m}+{4\pi^{2}J^{2}c\left({\textbf{k}\cdot\textbf{m}}\right)^{2} \over
{\left(C/c\right)\omega_n^{2}+Jc k^{2}+\alpha\left|\omega_{n}\right|k^{2}}}\right]\\ \nonumber &=& \sum_{\textbf{m}}\exp\left[\sum_{\textbf{k},\omega}-4\pi^{2} J
{Jc\left|\textbf{k}\times\textbf{m}\right|^{2}\over {\left(C/c\right)\omega_n^{2}+Jc k^{2}+\alpha\left|\omega_{n}\right|k^{2}}}- 4\pi^{2}J{\omega_{n}^{2}\textbf{m}\cdot\textbf{m}\left(C/c+\alpha
k^{2}/\left|\omega_{n}\right|\right)\over {\left(C/c\right)\omega_n^{2}+Jc k^{2}+\alpha\left|\omega_{n}\right|k^{2}}}\right],
\end{eqnarray}
\end{widetext}

\noindent where we have redefined $\alpha\rightarrow \alpha a^{3}$. The action in Eqn.\ref{dissvrt} has two possible phase transitions, depending on whether  the capacitance $C$ or the dissipation term $\propto \alpha$  in the numerator of the second term in Eq. (\ref{dissvrt})
dominates in long wavelength and low frequency limit. The former corresponds a critical point with the dynamic critical $z=1$, i.e. we recover the loop gas model which belongs to the 3d xy universality class. The latter corresponds, as we will show to $z = \infty$, i.e.  a fixed point with local criticality. The two
terms have the same scaling form for $z=2$. In the rest of the paper we analyze the dissipation driven transition

The vector field $\textbf{m}$ can be written as a sum of a
longitudinal, $\textbf{m}_{l}$, and a transverse component,
$\textbf{m}_{t}$ which by definition satisfy
$\overrightarrow{\nabla}\times\textbf{m}_{l}=0$ and
$\overrightarrow{\nabla}\cdot\textbf{m}_{t}=0$. The vorticity field $\rho_v\left(\textbf{r},\tau\right)$ is related to the transverse component alone and is given
by

\begin{equation}\label{vortex}
\overrightarrow{\nabla}\times
\textbf{m}_{t}\left(\textbf{r},\tau\right) = \rho_v\left(\textbf{r},\tau\right) = 
\rho_{vc}\sum_i\delta\left(\textbf{r}-\textbf{r}_{i}\right)\delta\left(\tau-\tau_{i}\right),
\end{equation}

\noindent where $\left(\textbf{r}_{i},\tau_{i}\right)$ are the
location of the core of the vortices and $\rho_{vc}$ is the quantized charge of the vortex.

Unlike the $z=1$ theory where
$d\textbf{m}/d\tau$ is the two component current that acts as the
additional topological defect, we find that the topological excitation in terms of which the problem with dissipation is most conveniently discussed
 involves only the longitudinal component and is given by a "warp" field $\rho_{w}\left(\textbf{r},\tau\right)$ defined through a non-local relation in space,

\begin{equation}\label{warprealspace}
{d\textbf{m}_{l}\left(\textbf{r},\tau\right)\over {d\tau}}=  \int d^2\textbf{R}  \rho_{w}\left(\textbf{R},\tau\right) {\textbf{r}-\textbf{R}\over {\left|\textbf{r}-\textbf{R}\right|^{3}}}.
\end{equation}
\noindent

\begin{figure}[h]
  \begin{center}
  \includegraphics[width=0.8\columnwidth]{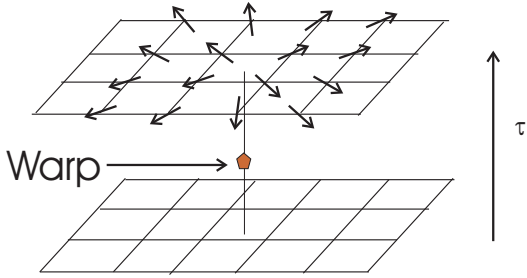}
  \caption{A "warp" is an event in time where the divergence of $\textbf{m}$ field changes}
  \label{fig:warp}
  \end{center}
\end{figure}

$\rho_{w}$ are events where the longitudinal component changes (see Fig.\ref{fig:warp}).
The warp charge $\rho_{wc}$ is defined by 
\begin{equation}\label{warpcharge}
\rho_{w}\left(\textbf{R},\tau\right) = \frac{c}{2\pi}\rho_{wc} \sum_i \delta(\textbf{R}-\textbf{R}_i) \delta(\tau-\tau_i),
\end{equation}
in a way analogous to the definition of the vortex charge $\rho_{vc}$ above. We will show in Sec. (III) that the warp-charge is also quantized.

Given Eq.(\ref{warprealspace}), the relation between the Fourier-transformed warp field $\rho_{w}\left(\textbf{k},\omega_{n}\right)$ and the longitudinal component of the velocity field is
\begin{equation}\label{warp}
{\imath\omega_{n}\over {ck}}\textbf{k}\cdot\textbf{m}_{l}\left(\omega_{n},\tau\right) =\rho_{w}\left(\textbf{k},\omega_{n}\right)
\end{equation}

\noindent In terms of $\rho_v, \rho_w$, the action, in the continuum
limit, neatly splits into three parts: $S=S_{v} + S_{w} + S'_{w}$
\begin{eqnarray}\label{chgact}
S_{v} &=& {1\over {L^{2}\beta}}\sum
{J\over {k^{2}}}\left|\rho_{v} (\textbf{k}\omega_{n})\right|^{2}\\
\nonumber S_{w} &=& {1\over {L^{2}\beta}}\sum{\alpha\over
{4\pi\left|\omega\right|}}\left|\rho_{w}(\textbf{k}\omega_{n})\right|^{2}
\\ \nonumber S'_{w} &=& {1\over {L^{2}\beta}}\sum
G\left(\textbf{k},\omega_{n}\right) \left(J J_{t} - {\alpha
J_{t}\left|\omega_{n}\right|\over {4\pi c}} - {\alpha^{2}k^{2}\over
{16\pi^{2}}}\right)\\ \nonumber && \times
\left|\rho_{w}(\textbf{k}\omega_{n})\right|^{2}.
\end{eqnarray}

\noindent where

\begin{equation}
G\left(\textbf{k},\omega_{n}\right) =  {1\over{J c k^{2} +
\left(C/c\right)\omega_{n}^{2}+\alpha\left|\omega_{n}\right|k^{2}}}
\end{equation}

\noindent The factorization of the action in eqn.\ref{chgact} is only possible due to the choice of variables made in eqn.\ref{warp}. With this choice, it follows from Eq.(\ref{chgact}) there are logarithmically interacting vortex charges in space with local interactions in time, and a logarithmically interacting warps in time with local interactions in space. The former is the standard Kosterlitz-Thouless-Berezinskii
theory \cite{KT,BER} while the latter is the charge representation of the Kondo problem \cite{AHY} near the Ferromagnet/Anti-Ferromagnet quantum critical point. Quite unlike the $z=1$ theory
where Lorentz invariance is manifest, the space and time degrees of freedom are completely decoupled in the theory describing the dissipation driven transition. Being orthogonal the charges
$\rho_v$ and $\rho_w$ are uncoupled; the action is a product of the action over configurations of $\rho_v$ and of $\rho_w$. Any physical correlations are determined by correlations of both
charges. The third piece of the action $S'_{w}$ is nonsingular because it corresponds to the three dimensional Coloumb gas problem which is known to be disordered at all temperatures \cite{POL}.

The singularity induced by the presence of Ohmic dissipation is transparent in our formulation. Suppose, one works instead  with the degrees of freedom appropriate to the
 2+1 d xy model which are the vortex density $\rho_{v}$ and the vortex current, $\textbf{J}_{v}$. The latter is defined as

\begin{equation}
\textbf{J}_{v} = \hat{z}\times {d\textbf{m}\over {d\tau}}
\end{equation}

\noindent The action in eqn.\ref{dissvrt} is

\begin{widetext}
\begin{eqnarray}\label{3dxydissvrt}
Z = \sum_{\textbf{m}}\exp\left[\sum_{\textbf{k},\omega} \right. &-& \left. 4\pi^{2} J {Jc \left|\rho_{v}\left(\textbf{k}\omega\right)\right|^{2}\over {\left(C/c\right)\omega_n^{2}+Jc
k^{2}+\alpha\left|\omega_{n}\right|k^{2}}}- 4\pi^{2}J{{\textbf{J}_{v}\cdot\textbf{J}_{v}}\left(C/c\right)\over {\left(C/c\right)\omega_n^{2}+Jc k^{2}+\alpha\left|\omega_{n}\right|k^{2}}}\right. \nonumber \\
&-& \left. 4\pi^{2}J{{\textbf{J}_{v}\cdot\textbf{J}_{v}}\left(\alpha k^{2}/\left|\omega_{n}\right|\right)\over {\left(C/c\right)\omega_n^{2}+Jc k^{2}+\alpha\left|\omega_{n}\right|k^{2}}}\right],
\end{eqnarray}
\end{widetext}

\noindent The first two terms can be combined to give the dual representation of the $3d xy$ model. In the absence of dissipation, these terms lead to the superconductor to insulator phase
transition. The third term introduces a new singularity in the vortex currents. Due to the continuity equation between $\rho_v$ and ${\bf J}_v$, see below, the two singularities cannot be decoupled as before and the analysis of the
model is not straightforward.

In terms of warps and vortices the vector field can be written as,

\begin{equation}
\textbf{m} = \rho_{v}\hat{z}\times \imath\textbf{k}/k^{2} + \rho_{w}c\hat{k}/\imath\omega
\end{equation}

\noindent and

\begin{equation}
\textbf{J}_{v} = -\rho_{v} \omega \textbf{k}/k^{2} + \rho_{w}\hat{z}\times c\hat{k}
\end{equation}

\noindent Separating the longitudinal and transverse components of the vector field $\textbf{m}$ and the vector current $\textbf{J}_{v}$ allows us to factorize the action. Notice that the new
term in the vector current is transverse and the continuity equation. $d\rho_{v}/d\tau +\nabla\cdot \textbf{J}_{v}=0$ is satisfied. 

Since warps are excitation that introduce longitudinal components in the vector field, alternate formulations could have been considered. In particular, we could have chosen to define monopoles
(sources and sinks in two dimensions) which are of the form

\begin{equation}\label{monopoles}
\overrightarrow{\nabla}\cdot\textbf{m}=\rho_{m}
\end{equation}

\noindent Such excitations determine the longitudinal component of the vector field and their relation to warps in momentum space is,

\begin{equation}\label{warp_monopole}
\rho_{m}\left(\omega,\textbf{k}\right) ={ck \over \omega} \rho_{w}\left(\omega,\textbf{k}\right).
\end{equation}

\noindent In terms of monopoles one cannot separate the singularities and the simple description of the dissipation driven transition is no longer possible. The physics remains same as the
scaling dimension of monopoles and warps is the same. We choose to work with warps as the analysis is more transparent and we do not have to consider the added complication of the
renormalization of velocity.

\section{Warps as topological defects in space time}

In this section we discuss the topological properties and the physical content of the warps. The definitions in Eqs. (\ref{warp}) and (\ref{warprealspace}) say  that the warps are source for the longitudinal component of the vector field $\textbf{m}$. Since $\textbf{m}$ is an integer vector field, it is important also to show that this property is preserved by Eq. (\ref{warprealspace}) as well as that the warp-charge $\rho_{wc}$ is quantized.  

Let us first review the familiar case of the vortex in Eqn.\ref{vortex}. In real space the vector field $\textbf{m}_{t}$  due to a vortex at $\textbf{R}_{i}$ is

\begin{equation}
\label{vortex}
\textbf{m}_t\left(\textbf{r},\tau\right) = \rho_{vc} {\widehat{z} \times \left(\textbf{r}-\textbf{R}_{i}\right)\over {\left|\textbf{r}-\textbf{R}_{i}\right|^{2}}}
\end{equation}

\noindent Given the integer vector field $\textbf{m}$  the vortex is defined by its quantized charge $\rho_{vc}$ and the singular azimuthal vector field $\propto 1/r$. This vector field is undefined inside a core radius $a$ where the source of the velocity field with discrete charge $\rho_{vc}$ sits. The discreteness of $\textbf{m}_{t}$ is respected once such a singularity is identified. In particular, the corresponding $\theta$ field has a discontinuity along a line going from the core out to infinity. One can then proceed to the continuum to analyze the critical properties of the model . 

We see from the solution of Eq. (\ref{warprealspace}) that the field ${\bf m}_{l}$  due to a  warp created at site $\textbf{R}_i$ and time $\tau_i$ has the form:
\be
\label{warp1}
{\bf m}_{l}({\bf r},\tau) = c\rho_{wc} \Theta(\tau-\tau_i) \frac{{\bf r- R_i}}{|{\bf r}-{\bf R}_i|^3}
\ee
\noindent Analogous to the vortex, this solution is a quantized charge in a core which serves as the source of  a radial velocity field falling off as $1/r^{2}$. To see this consider the solution to eq.\ref{warp} with $\textbf{m} = \nabla \theta$,

\begin{eqnarray}\label{theta_warp}
\theta\left(\textbf{r},\tau\right) &=& \Theta\left(\tau-\tau_{i}\right)\left(2\pi n {a\over \left|\textbf{r}-\textbf{R}_{i}\right|} + \theta_{c}\right), \left|\textbf{r}-\textbf{R}_{i}\right| \geq a \nonumber \\
&=& \Theta\left(\tau-\tau_{i}\right) \theta_{c}, \left|\textbf{r}-\textbf{R}_{i}\right| < a
\end{eqnarray}

\noindent The boundary conditions chosen are $\theta(\left|\textbf{r}-\textbf{R}_{i}\right| = a) = \theta_{c} + 2\pi n$ and $\theta\left(\left|\textbf{r}-\textbf{R}_{i}\right| = \infty\right) = \theta_{c}$. The first guarantees the quantization of the warp charge, imposed by the phase jump of $2\pi n$ that occurs at $\left|\textbf{r}-\textbf{R}_{i}\right| = a$. In other words a  warp is an event which produces a discrete phase jump across the core. The corresponding discontinuity in the phase is discrete, corresponding to a discrete $\textbf{m}_{\ell}$ field localized along a loop encircling the core. This localized discrete $\textbf{m}_{\ell}$ serves as the source of the radial ${\bf m}_{\ell}$ field outside the core. As one moves along any radial direction crossing this loop, which defines the boundary of the core, the phase jumps by $2\pi n$. Thus the warp charge is discrete and satisfies the constraint that the vector field $\textbf{m}$ is discrete. The second boundary condition where is chosen because we are only describing the changes with respect to the fields because of the warp event at $\tau =\tau_i$.

\section{Dissipation driven phase transition}

The action for the warps, Eq. (\ref{chgact}), has been studied before
in the context of quantum coherence of two state system coupled to
dissipative baths \cite{BM,SC3}. Let us introduce a core-energy
$\Delta$ for the $\rho_w$'s just as is done to control the fugacity
of $\rho_v$'s, the vortices. Next consider how the renormalization
of $\alpha$ and $\Delta$ proceeds. \cite{footnote}. Including the core-energy, the
action $S_w$ is
\begin{equation}
S_w = \sum_{i}\left[T\sum_{n}{\alpha\over {4\pi}}{1\over
\left|\omega_{n}\right|}\left|\rho_{wi}\left(\omega_{n}\right)\right|^{2}
+ \int d \tau
\Delta\left|\rho_{wi}\left(\tau\right)\right|^{2}\right].
\label{warpaction}
\end{equation}

\noindent The RG equations for eqn.\ref{warpaction} are well known
\cite{AHY,BM,SC3,LEG}

\begin{eqnarray}\label{rgflow}
{d \widetilde{\Delta}\over {dl}} &=& \left(1 - \alpha\right)\widetilde{\Delta}\\
\nonumber {d\alpha \over {dl}} &=& -\alpha \widetilde{\Delta}^{2},
\end{eqnarray}

\noindent where $\widetilde{\Delta} = \Delta\tau_{c}$ and $\tau_{c}
\approx (2 J_2E)^{-1/2}$ is the short time cutoff. The critical
point of interest is at $\alpha_{c} = 1$, where $\widetilde{\Delta}$
scales to 0; for $\alpha <1$ the charges $\rho_w$ freely proliferate
as "screening" due to $\widetilde{\Delta}$ becomes effective.
$\alpha >1$ represents the ordered or confined region. We are
interested here only in the region $\alpha \leq 1$. Well in the
quantum critical region, the (singular part of the) propagator for
$\rho_w$ is

\be \label{rho0corr}
\left<\rho_{w}\left(\omega_{n}\right)\rho_{w}\left(-\omega_{n}\right)\right>
= \left(1/4\pi\left|\omega_{n }\right|\tau_{c}\right)^{-1}.
\ee

\begin{figure}[h]
  \begin{center}
  \includegraphics[width=0.8\columnwidth]{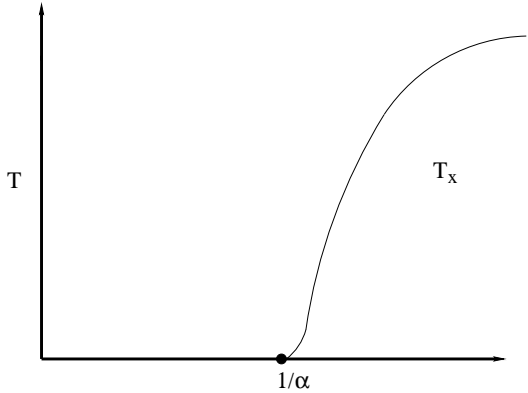}
  \caption{Dissipative driven proliferation of warps occurs for $1/\alpha >1$.
  The power law scaling at  criticality is cut off at an energy scale $T_{x}$ as discussed in the text.}
  \label{fig:talpha}
  \end{center}
\end{figure}

\noindent Note that the form of the propagator in eqn.\ref{rho0corr}
is valid only at the critical point. The crossover to the
quantum-disordered or screened state is given for $T=0$ when $\omega
= \omega_x$ which is of the order of the inverse of the
characteristic screening time, which may be estimated similarly to
Kosterlitz's estimate \cite{Kosterlitz} of the screening length in
the $xy$-problem:
 \be \label{crossover} \omega_x \approx
\tau_c^{-1} \exp\left(- b/\sqrt{1-\alpha}\right), \ee where b is a
numerical constant of O(1). At finite temperatures and low
frequencies, the crossover temperature $T_x$ is of  O($\omega_x$).
On the ordered side $\alpha > \alpha_{c}$, we have to consider the
effect of the charging energy and the flow to the $z=1$ critical
point. In fig.\ref{fig:talpha} we plot the phase diagram in the
$T-\alpha$ plane where only the the region $\alpha\geq \alpha_{c}$
is correctly described. The crossover line is also shown where well
above $T_{x}$ we expect to observe critical behavior controlled by
the quantum critical point while below $T_{x}$ the system is phase
disordered and hence an insulator.

\section{Correlation Function}

We can compute the correlation function for the order parameter field following a procedure which is a generalization of the method developed for the classical 2DXY model \cite{JOS}. We are
interested in the expectation value of $\exp\left(\imath\theta_{i\mu}\right)$ given by

\begin{equation}
\label{orderparameter} C_{i,j,\mu,\nu} = \left<e^{\imath
\theta_{i\mu}}e^{-\imath\theta_{j\nu}}\right> \equiv { \int
d\left[\theta\right]e^{-\bar{S}}\over {\int
d\left[\theta\right]e^{-S}}},
\end{equation}

\begin{equation}
C = \left<e^{\imath\theta_{i\mu}-\imath\theta_{j\nu}-2
\pi\imath\sum_{path}\textbf{m}}\right>
\end{equation}

\noindent Notice that the exponential of the sum along the path  is
always one as $\textbf{m}$ is an integer fields. We need to include it in our calculations to reproduce the correlation functions in the absence of dissipation.  Since
$\textbf{m}$ is two dimensional vector, the contribution is only
from the space like segment of the path chosen. We can now proceed
as before. Going over to imaginary time implies that the factor of
$\imath$ is absorbed in the definition of the path integral.

\begin{figure}[h]
  \begin{center}
  \includegraphics[width=0.8\columnwidth]{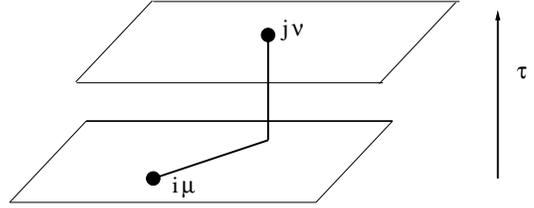}
  \caption{Path between $i\mu$ and $j\nu$ used to compute the correlation function.}
  \label{fig:path}
  \end{center}
\end{figure}

\noindent To compute the correlation function we introduce a field
$\overrightarrow{\eta}_{i\mu}$ that lives on the sites of the space
time lattice. Its $x$ and $y$ component are $1$ or $0$ depending on
whether or not the path includes the link $\left(xy\mu\right)$ to
$\left(x+1y\mu\right)$ or $\left(xy+1\mu\right)$ respectively.
Similarly we introduce a field $\eta^{0}_{i\mu}$ which is $1$ or $0$
depending on whether or not the path includes the link
$\left(i\mu\right)$ to $\left(i\mu + 1\right)$. The path we choose
to compute the correlation function is shown in fig.\ref{fig:path}.
We can now integrate out the phase degrees of freedom as before to
obtain the functional in terms of the vector fields alone.

\begin{eqnarray}
\bar{S} = S _v &+& S_w  \nonumber \\ &+& {1\over
{L^{2}\beta}}\sum_{ln}\left(2\pi\right)J
G\left(\textbf{k}_{l},\omega_{n}\right)\left[\left(-\imath\omega_{n}\eta^{0\star}/2\right)\left(\imath\textbf{k}_{l}\cdot\textbf{m}_{ln}\right)
 \nonumber \right. \\  &\ &  +
\left(-\imath\textbf{k}_{l}\times\overrightarrow{\eta}^{\star}\right)\cdot\left(\imath\textbf{k}_{l}\times\textbf{m}_{ln}\right)
+c.c.  \nonumber \\ &\ & + \left(\left(C/c\right)\omega_{n}^{2}+\alpha\left|\omega_{n}\right|k^{2}\right)\left(\textbf{m}\cdot\overrightarrow{\eta}\right)\nonumber \\
&\ & \left. +
 \left|\omega_{n}\eta^{0}/2+\textbf{k}_{l}\cdot\overrightarrow{\eta}^{\star}/2\right|^{2}\right].
\end{eqnarray}

\noindent Given the definition of vortex and warps, we can write the
vector field as

\begin{equation}\label{vecfld}
\textbf{m} = \rho_{v}{\imath\textbf{k} \times \widehat{z}\over
{k^{2}}} + {\rho_{w}{c\over {\imath\omega}}}\widehat{k}
\end{equation}

\noindent Replacing in the expression above we see that the vortex
part of the correlation simplifies just as it did in the case of the
action itself.

\begin{eqnarray}\label{modcorr}
\bar{S} = S _v &+& S_w  \nonumber \\ &+& {1\over
{L^{2}\beta}}\sum_{ln}\left(2\pi\right)J
G\left(\textbf{k}_{l},\omega_{n}\right)\left[\left\{\left(-\imath\omega_{n}\eta^{0\star}/2\right){k\over
\omega}c\rho_{w}
 \nonumber \right. \right.\\  &\ & \left.  + {\left(\left(C\right)+c \alpha k^{2}/\left|\omega\right|\right)\over k_{l}}\rho_{w}\left(\textbf{k}\cdot\overrightarrow{\eta}\right)+c.c.\right\}\nonumber \\
&\ & \left. +
 \left|\omega_{n}\eta^{0}/2+\textbf{k}_{l}\cdot\overrightarrow{\eta}^{\star}/2\right|^{2}\right]\nonumber
 \\ &\ & +\left\{ {2 \pi J\over
 {k^{2}}}\left(-\imath\textbf{k}_{l}\times\overrightarrow{\eta}^{\star}\right)\cdot\rho_{v}+
 cc. \right\}
\end{eqnarray}

\noindent The net correlation is a product of three distinct
contributions. The term independent of vortices and warps is the spin wave contribution.
Notice that the associated propagator is three dimensional and as
such is non singular. This reproduces the well known result that
spin waves do not disorder the long range order in three dimensions.
The other two contributions come from vortices (the last line in
eqn.\ref{modcorr}) and warps. At the dissipative critical point, the
singular fluctuations are associated with the warps. Thus the
vortices and spin wave contributions are completely regular. To
compute the contribution from the warps we follow a procedure
similar to Jose et al. (\cite{JOS}). Consider the contribution from
the warps,

\begin{eqnarray}
\bar{S_{w}} &=& S_{w} + {1\over
{L^{2}\beta}}\sum_{ln}\left(2\pi\right)J
G\left(\textbf{k}_{l},\omega_{n}\right)\left[\left\{\left(-\imath\omega_{n}\eta^{0\star}/2\right){ck\over
\omega}\rho_{w}\right.\right. \nonumber \\ &\ & \left. \left. +
{\left(C + \alpha c k^{2}/\left|\omega\right|\right)\over
k_{l}}\rho_{w}\left(\textbf{k}\cdot\overrightarrow{\eta}\right)+c.c.\right\}\right]
\end{eqnarray}

In real space the terms of the action can be rewritten as

\begin{eqnarray}\label{modcorrrs}
\bar{S_{w}} &=& S_{w} +  \int  d\textbf{R}d\textbf{R}'d\tau d\tau '
\left[\partial_{\textbf{r}}\overrightarrow{\eta}\left(\textbf{R},\tau\right)\right.\\
&\ & G \left(\textbf{R}\tau;\textbf{R}'\tau '\right)\left[{\left(C +
\alpha c k^{2}/\left|\omega\right|\right)\over
\imath k_{l}}\rho_{w}\right]\left(\textbf{R}'\tau '\right) \nonumber \\
\nonumber &+& \left.
\partial_{\tau}\eta^{0}\left(\textbf{R},\tau\right) G
\left(\textbf{R}\tau;\textbf{R}'\tau '\right)\left[{ck \over
{\omega}}\rho_{w}\right]\left(\textbf{R}'\tau '\right)\right],
\end{eqnarray}

\noindent where the derivatives are taken along the path and the
term $[..]\left(\textbf{R},\tau\right)$ is meant to represent the
fourier transform of the terms within the square brackets. Consider
the derivatives along the path of the fields $\overrightarrow{\eta}$
and $\eta^{0}$. They represent the change in value of the fields
computed from one point to the next along the chosen path. Since the
first term in eqn.\ref{modcorrrs} couples only to the spatial
derivative it has finite values only at
$\left(\textbf{r},\tau\right)$ and $\left(\textbf{r}',\tau\right)$
(ends of the space like component of the path) while the second term
couples to time derivatives which are finite at
$\left(\textbf{r}',\tau\right)$ and $\left(\textbf{r}',\tau
'\right)$.

\begin{eqnarray}
\bar{S_{w}} &=& S_{w} + \int d\textbf{R}d\tau_{1}\left[{ck
\over \omega}\rho_{w}\right]\left(\textbf{R}\tau_{1}\right)\nonumber \\
\nonumber &\ &
\left\{G\left(\textbf{r}'\tau,\textbf{R}\tau_{1}\right)-G\left(\textbf{r}'\tau
',\textbf{R}\tau_{1}\right)\right\}\\ \nonumber &+& \int
d\textbf{R}d\tau_{1}\left[{\left(C + \alpha c
k^{2}/\left|\omega_{n}\right|\right)\over
k_{l}}\rho_{w}\right]\left(\textbf{R}\tau_{1}\right)\nonumber \\
&\ &
\left\{G\left(\textbf{r}\tau,\textbf{R}\tau_{1}\right)-G\left(\textbf{r}'\tau
,\textbf{R}\tau_{1}\right)\right\}
\end{eqnarray}

\noindent Fourier transforming back we get

\begin{eqnarray}
\bar{S} = S _v &+& S_w  \\  &+& {1\over
{L^{2}\beta}}\sum_{ln}\left(2\pi\right){J G\over
2}\left(\textbf{k}_{l},\omega_{n}\right)\left[\left(
e^{\imath\textbf{k}_{l}\cdot\left(\textbf{r}_{j}\right)+\imath\omega_{n}\left(\tau_{\mu}\right)}
\right. \right. \nonumber \\  &-& \left.\left.
e^{\imath\textbf{k}_{l}\cdot\left(\textbf{r}_{j}\right)+\imath\omega_{n}\left(\tau_{\nu}\right)}\right)\left({ck
\over {\omega}}\rho_{w}\right) + \left(
e^{\imath\textbf{k}_{l}\cdot\left(\textbf{r}_{i}\right)+\imath\omega_{n}\left(\tau_{\mu}\right)}\right.\right.\nonumber\\
\nonumber &-& \left.\left.
e^{\imath\textbf{k}_{l}\cdot\left(\textbf{r}_{j}\right)+\imath\omega_{n}\left(\tau_{\mu}\right)}\right)\left({\left(C
+ \alpha c k^{2}/\left|\omega_{n}\right|\right)\over
k_{l}}\rho_{w}\right)\right]
\end{eqnarray}

\noindent Since the action $S_{w}$ is quadratic, we can compute the
correlation to give
\begin{eqnarray}\label{gv}
&&C({\bf r-r'}, \tau-\tau') \propto \exp (-F), \\
\nonumber &&F = -{ J^{2}c^{2}\over 4}T\sum_{n}\int
{d\textbf{k}}\left<\rho_{w}\left(\omega_{n}\right)\rho_{w}\left(-\omega_{n}\right)\right>\times  \\
\nonumber &&
G^{2}\left(\textbf{k},\omega_{n}\right)\left[{k^{2}\over
{\omega_{n}^{2}}}\left(1-
\cos\left(\omega_{n}\left(\tau-\tau'\right)\right) \right)\right.\\
\nonumber && +\left({\left(\left(C/c\right) + \alpha
k^{2}/\left|\omega_{n}\right|\right)\over k_{l}}\right)^{2}\left(1-
\cos\left(\textbf{k}_{l}\cdot\left(\textbf{r}_{i}-\textbf{r}_{j}\right)\right)
\right)\\ \nonumber && +\left({\left(\left(C/c\right) + \alpha
k^{2}/\left|\omega_{n}\right|\right)\over
\omega_{n}}\right)\left(e^{\imath\textbf{k}_{l}\cdot\left(\textbf{r}_{i}-\textbf{r}_{j}\right)}+e^{\imath\omega_{n}\left(\tau_{\mu}-\tau_{\nu}\right)}\right.\\
\nonumber && \left.\left.-
e^{\imath\textbf{k}_{l}\cdot\left(\textbf{r}_{i}-\textbf{r}_{j}\right)+\imath\omega_{n}\left(\tau_{\mu}-\tau_{\nu}\right)}-1\right)\right]
\end{eqnarray}

\noindent The order parameter correlation function has three
contributions from the warps. The first term in the square brackets
in eqn.\ref{gv} is from the time like segment of the path; the
second term corresponds to the space like segment of the path and
the last  term is an interference term between the two. At the
critical point the correlation of the warps is given by
eqn.\ref{rho0corr}. Consider the space like segment. It is singular
(of the form $\left(1/\left|\omega_{n}\right|\right)$) as $\omega$
goes to zero.  This implies $F\rightarrow \infty$ and the
correlation function is zero unless $\textbf{r}_{i}=\textbf{r}_{j}$.
Thus the correlation function is local. For
$\textbf{r}_{i}=\textbf{r}_{j}$ the last two terms are zero and only
the time like path contributes to the correlation function.
Performing the integral over momentum $\textbf{k}$, the correlation
function is given by $C(0,\tau-\tau') \propto
\exp(-F(0,\tau-\tau'))$, where
\begin{equation}\label{gv1}
 F(0,\tau-\tau')=-{ 2\pi T}\sum_{n} {1-
\cos\left(\omega_{n}\left(\tau-\tau'\right)\right)\over
\left|\omega_{n}\right|}\log\left(
{\left|\omega_{n}\right|\tau_{c}}\right).
\end{equation}

\noindent To perform the matsubara sum we consider the analytic
properties of the logarithm in eqn.\ref{gv1}. The branch cut
introduced on the real axis implies that on integrating over a
contour infinitesimally above the real axis from $-\infty$ to
$\infty$ and infinitesimally below the real axis from $\infty$ to
$-\infty$ the real part cancel while the imaginary part enforces a
sum over positive frequencies giving

\begin{equation}\label{gv2}
 F(0,\tau-\tau')=-{ 2\pi T}\sum_{n=1}^{\infty} {1-
\cos\left(\omega_{n}\left(\tau-\tau'\right)\right)\over
\left|\omega_{n}\right|}.
\end{equation}

\noindent Such correlation functions have been calculated in other
contexts. In particular using the calculation by Ghaemi et al.
\cite{GAS}, the spectral function of the correlation is

\begin{equation} \label{spectra}
Im \chi({\bf q},\omega) =
\bar{\tau}_c\tanh(\omega/2T),~~\omega \lesssim \tau_c^{-1}.
\end{equation}

\noindent $\bar{\tau}_c$ is $O(\tau_c)$. The order parameter correlation at $\alpha=1$ is local in space and power law in time. Thus the dissipative 2DXY model describes a new type of quantum
criticality where the correlations are local in space and are functions of $\omega/T$, with a cut-off provided above. Precisely such criticality had been postulated to describe the "Strange metal" or marginal fermi-liquid region of the phase diagram of high temperature superconductors \cite{mfl}.

\section{Zero Temperature Phase Diagram}

So far we have concentrated on the dissipation driven transition. We now consider the phase diagram in the $JC - \alpha$ plane. Our analysis clearly shows that the physics at finite
dissipation and at zero dissipation cannot be continuously connected. This is in accord with the suggestion in  Ref.( \onlinecite{SC1}), that the zero dissipation limit, described by the $3d-xy$ model at zero temperature, is in a different
universality class compared to the model at infinitesimal dissipation. The other limit where the physics is easy to analyze is at infinite
dissipation. In this case, the fluctuations in time are energetically unfavored and the model maps to a $2D-xy$ model. We now make a guess on the structure of the rest of the phase diagram based on the well-defined theory in the limits discussed.

In reference [6] it is argued that a weak coupling analysis shows that the entire Josephson coupling renormalizes to zero at small dissipation. Furthermore a self consistent calculation allows one
to map the phase boundary within this scheme. Such an analysis does not capture the physics of topological defects that are necessary to determine the entire phase diagram. Given our
identification of the relevant degrees of freedom and the fugacities associated with the new excitations, we propose an alternative understanding of the physics of the dissipative $2d-xy$ model.

\begin{figure}[h]
  \begin{center}
  \includegraphics[width=0.8\columnwidth]{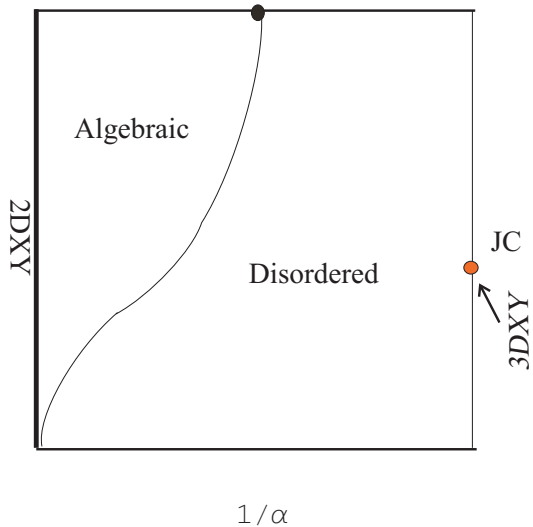}
  \caption{The zero temperature phase diagram in the $1/\alpha-JC$ plane for the dissipative 2DXY model. For $\alpha=0$, the system undergoes a superconductor to insulator phase transition with a 3DXY critical point.
  For infinite $\alpha$, the dynamics is quenched and the we recover the 2DXY behavior. There are two phases for finite dissipation. The disordered phase is characterized with a diverging fugacity of
  warps and vortices. The algebraically ordered phase is one in which both the vortices and warps are bound, and the correlations are power laws separately in space and time with continuously varying
  exponents. }
  \label{fig:phsdiag}
  \end{center}
\end{figure}

We have explicitly shown above that the critical theory has two independent singularities. Thus the true phase diagram can be determined by studying the flow of stiffness, fugacity of vortices
and fugacity of warps. The only algebraically ordered phase is one in which both the vortices and warps are bound. This occurs for $1/\alpha<1$ and $JC>2/\pi$. Furthermore this phase is
continuously connected to the 2d XY phase at infinite dissipation. Thus one expects a phase boundary to exist that starts at $JC=\infty$ and $\alpha=1$ and approaches$JC=0$ and
$\alpha=\infty$ asymptotically as shown in fig.\ref{fig:phsdiag}. For $1/\alpha>1$ the fugacity of warps grows and the stiffness vanishes. Similarly, for $JC<2/\pi$ the vortices proliferate
driving the stiffness to zero. We find no evidence for the ordered phase for small $JC$ and $\alpha>1$ as in reference [6] where the large fugacity of the vortices was ignored. Similarly, for large $JC$ a minimum value of $\alpha$ is required to overcome the fugacity of the warps. In other words for any given value of $JC$ as one increases
$\alpha$ the fugacity of warps, which is infinite at infinitesimal $\alpha$, remains large until we cross the phase boundary, upon which it renormalizes to zero. Whether the resulting phase is
algebraically ordered or not depends on the fugacity of vortices.

Consider the regime where warps proliferate. The vanishing stiffness implies that such excitations should also lead to screening in the interaction of the vortices. In analogy with the
disordered phase in the KT theory, we should introduce a mass for both vortices and warps. This in turn will lead to unbinding of vortices as well. We believe that for small $JC$, this
proliferation of warps and resultant screening of interaction among vortices will push the phase boundary away from $\alpha=1$ to larger values of $\alpha$. At large $JC$ we expect our theory to
be valid and a dissipation driven transition will occur at $\alpha=1$. A schematic phase diagram in shown in fig.\ref{fig:phsdiag}.

At first sight it might appear that a long range interaction in time of the form $1/\left|\tau-\tau'\right|^{2}$ is inconsistent with a disordered ground state, where the correlation function
decays faster than the interaction. The reason such a state is allowed is the fact that the existence of excitations which correspond to phase slip events in time introduces a second time scale
in the problem. The competition between the short time scale physics associated with the fugacity of the warps and the long range interaction between them is precisely the reason for the
observed phase transition.

\section{Ashkin Teller model and the Cuprates}

In reference [1] we have argued that the long wavelength theory of a model for the Pseudogap state for the Cuprate superconductors is the Ashkin-Teller model, The quantum version of this model with dissipation  is in the universality class of dissipative quantum $2d-xy$
model. The Marginal Fermi Liquid \cite{mfl} properties of the optimally doped system originate from the fluctuation spectrum of the local quantum critical point. In this model the phase
variable $\theta$ represents the possible current loop order within a given unit cell and the four fold anisotropy restricts it to four values representing states that break time reversal and
one of two possible reflection symmetries. The origin of the dissipation in this system is due to the coupling of the order parameter, which corresponds to the coherent part of the current
operator, to the incoherent part of the current operator which correspond to fermions near the Fermi surface. For long-wavelength fluctuations such a coupling leads to Ohmic dissipation of the
form in eqn.\ref{diss}, while term of the form in eqn.\ref{densdiss2}(see below) is not generated. Detailed derivation of the long wavelength theory will be provided in a later publication.

\begin{figure}[h]
  \begin{center}
  \includegraphics[width=0.8\columnwidth]{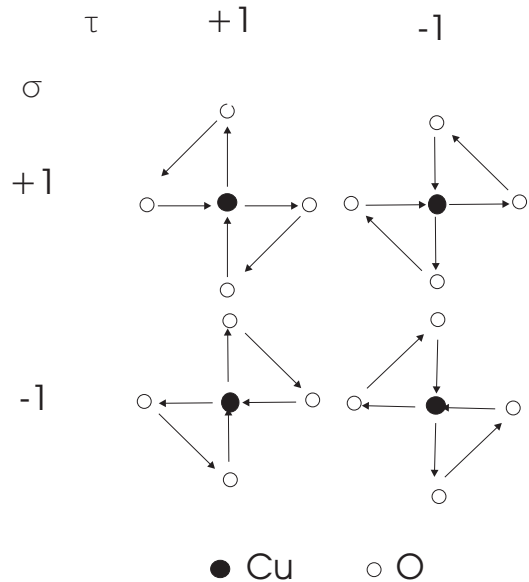}
  \caption{The four domains of the circulating current phase is shown. They can be parameterized by two Ising variables which represent the currents in the $x$ and $y$ directions.}
  \label{fig:domain}
  \end{center}
\end{figure}

The classical Ashkin Teller model is defined in terms of two Ising degrees of freedom per unit cell. If we define $\sigma_{i}$ and $\tau_{i}$ as Ising variables taking $\pm 1$ values to
parameterize the currents along the the $x$ and $y$ direction, the four resulting states in each unit-cell (see fig.\ref{fig:domain}) form the basis of the low energy sector of the long wavelength theory.

\noindent The Hamiltonian of the classical Ashkin Teller model is \cite{AT}

\begin{equation}\label{hat}
H_{AT} = -\sum_{<ij>}\left[J_{2}\left(\sigma_{i}\sigma_{j} + \tau_{i}\tau_{j}\right)+J_{4}\sigma_{i}\sigma_{j}\tau_{i}\tau_{j}\right]
\end{equation}

\noindent For $J_{4}=0$ the model reduces to two decoupled Ising models which is known to undergo a Gaussian phase transition \cite{IZ}. The analysis has been extended to map out the entire
phase diagram by observing that $J_{4}$ is a marginal operator in this theory \cite{KD}. For $-1<J_{4}/J_{2}<1$ the system undergoes a Gaussian phase transition form a ferromagnetically ordered
state to a paramagnetic state with continuously varying exponents. Quite remarkable, for $-1<J_{4}/J_{2}<0$, the specific heat exponent $\alpha$ is negative. The Ashkin Teller model
explains why no singularity is observed in specific heat measurements at the pseudogap temperature accompanying the breaking of time reversal symmetry. Defining
$\sigma_{i}=\cos\left(\theta_{i}\right)$ and $\tau_{i}=\sin\left(\theta_{i}\right)$ we can rewrite the Hamiltonian as

\begin{equation}\label{hatphase}
H_{AT} = -\sum_{<ij>}\left[2 J_{2} \cos\left(\theta_{i}-\theta_{j}\right)+J_{4}\cos\left(2\theta_{i}-2\theta_{j}\right)\right]
\end{equation}

\noindent where $\theta_{i}$'s take values $\pi/4,3\pi/4,5\pi/4$ and $7\pi/4$. The constraint can be implemented as a four fold anisotropy term,

\begin{eqnarray}\label{hath}
H_{AT} = &-&\sum_{<ij>}\left[2 J_{2} \cos\left(\theta_{i}-\theta_{j}\right)+J_{4}\cos\left(2\theta_{i}-2\theta_{j}\right)\right]\nonumber \\  &+& \sum_{i}h_{4}\cos\left(4\theta_{i}\right)
\end{eqnarray}

\noindent Given this mapping of the Ashkin Teller model to a xy model with four fold anisotropy, we now analyze the effect of $h_{4}$ on the local quantum critical point of the dissipative 2dxy
model.

\section{Effect of four fold anisotropy}

The discrete nature of the underlying degrees of freedom  of the Ashkin Teller model enforce an Ising type order at low temperatures. On increasing temperature, it undergoes a Gaussian
transition into a state which can be described by proliferation of vortices. It is a unique property of four fold anisotropy that the disordering, which occurs via proliferation of vortices,
and ordering, which is enforced by a diverging anisotropy field, happen at the same temperature \cite{JOS}. We now analyze the effect of such an anisotropy field on the local quantum critical
point. To do so we introduce the term $H_{4} = \sum_{i}h_{4}\cos\left(4 \theta_{i}\right)$ in the Hamiltonian. To handle such a term in the action we employ the following approximation scheme

\begin{equation}\label{anis}
e^{h_{4}\cos\left(4\theta_{i}\right)} \approx
\sum_{p_{i}}e^{\ln\left(y_{4}\right)p_{i}^{2}+\imath4p_{i}\theta_{i}},
\end{equation}

\noindent where $p_{i}$ is an integer field that lives at each site
and $y_{4}=h_{4}/2$. For large values of $y_{4}$ the approximation
is reliable as the sum will be dominated by the terms with
$p_{i}=0,\pm 1$. Eq.\ref{anis} introduces two new terms in the
action, one linear in the phase and the other independent of it. We
can proceed as before in integrating out the phase degrees of
freedom to get a new action which is of the form

\begin{eqnarray}
S= S_{v}+ S_{w} &+& {1\over {L^{2}\beta}}\sum G c^{2}
\left(4\pi^{2}J {k\over
{\omega}}\rho_{w}\left(\textbf{k},\omega\right)
p\left(-\textbf{k},-\omega\right)\right. \nonumber \\   &+& \left.
\left|p\left(\textbf{k},\omega\right)\right|^{2}\right) +
\ln\left(y_{4}\right)\left|p\left(\textbf{k},\omega\right)\right|^{2}.
\end{eqnarray}

\noindent Given that the coupling between the integer field $p$ and the phase variable is linear we can follow a self consistent procedure to calculate its effect on the dissipative fixed point.
In this context it is useful to recall that in the absence of dissipation, the disordered side is characterized by the anisotropy field going to zero, i.e. being marginally irrelevant.
Furthermore, in the absence of anisotropy the dissipative transition leads to a state which is characterized by the proliferation of warps. In other words, the most singular part of the action
is the interaction between warps in time. Since the action for the integer field, $p$, is quadratic, the cumulant expansion implies that the a new term in the action for warps appears of the
form

\begin{eqnarray}\label{aniscoup}
\widetilde{S}_{4} &=&{1\over {2 L^{2}\beta}}\sum  \left(4\pi^{2}J G
c^{2} {k\over
{\omega}}\right)^{2}\rho_{w}\left(\textbf{k},\omega\right)\rho_{w}\left(-\textbf{k},-\omega\right)\times\nonumber
\\
&
&\left<p\left(\textbf{k},\omega\right)p\left(-\textbf{k},-\omega\right)\right>.
\end{eqnarray}

\begin{figure}[h]
\includegraphics[width=\columnwidth]{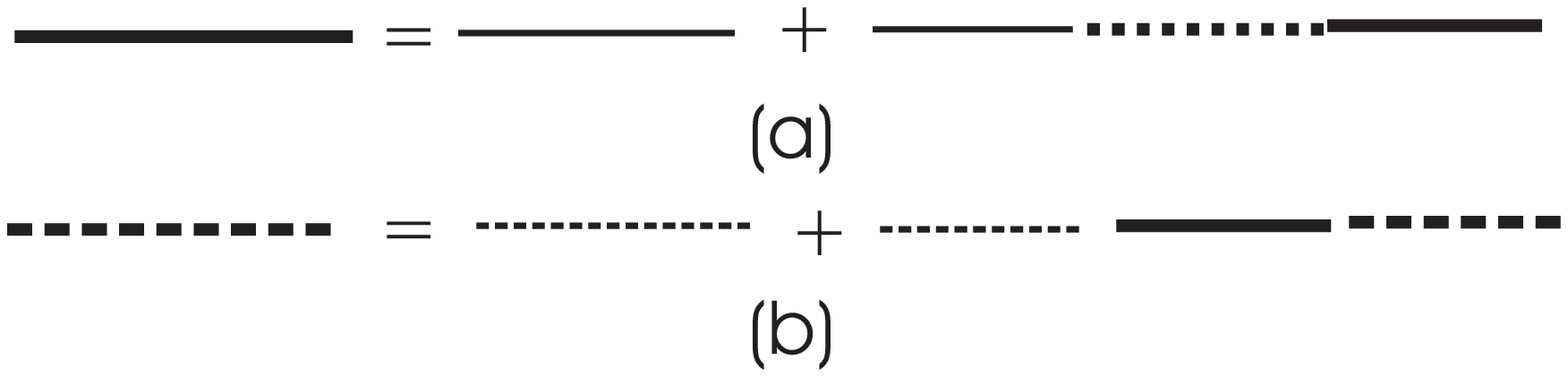}
  \caption{Equations for he full Greens function. The thick lines  represent the full propagator
  $\left<p\left(\textbf{k},\omega\right)p\left(-\textbf{k},-\omega\right)\right>$ and the thick dashed lines
  represent
  $\left<\rho_{w}\left(\textbf{k},\omega\right)\rho_{w}\left(-\textbf{k},-\omega\right)\right>$,
  while the thin full and dashed lines represent the same propagators computed in the absence of the linear coupling with the
  warps.}
  \label{fig:greensfn}
  \end{figure}

\noindent The correlation is obtained from a self consistent  solution of the coupled equations
shown diagrammatically represented in Fig.\ref{fig:greensfn}.

\noindent A quadratic equation is obtained for each of the
propagators. Since we are looking at the effect of the anisotropy on
the effective action for the warps, we can solve these equations to
obtain the most singular contribution to the propagator in
Eqn.\ref{aniscoup}. Defining
$G_{p}=\left<p\left(\textbf{k},\omega\right)p\left(-\textbf{k},-\omega\right)\right>$
and
$G_{w}=\left<\rho_{w}\left(\textbf{k},\omega\right)\rho_{w}\left(-\textbf{k},-\omega\right)\right>$,
symbolically the equation satisfied by $G_{p}$ can be written as

\begin{equation}
\lambda^{2}G_{w}^{0}G_{p}^{2} -G_{p}+G_{p}^{0} = 0.
\end{equation}

\noindent where $G_{p}^{0}$ and $G_{w}^{0}$ are the corresponding
propagators evaluated in the absence of the linear coupling and
$\lambda=\left(4\pi^{2}J G c^{2} {k/{\omega}}\right)$. At the
quantum critical point, the most singular contribution for the warps
is logarithmic in frequency, while the action for the integer fields
is not singular. This is again due to the fact that the integer
fields interact via Colomb Kernel in three dimensions. Thus the most
singular part of the solution is of the form $G_{p}\approx
1/\lambda^{2}\left[G_{w}^{0}\right]^{-1}$. Substituting in
Eqn.\ref{aniscoup}, the effect of the anisotropy is to introduce a
term in the action of the form

\begin{equation}
\widetilde{S}_{4}\approx {1\over {2 L^{2}\beta}}\sum
\left[G_{w}^{0}\right]^{-1}
\rho_{w}\left(\textbf{k},\omega\right)\rho_{w}\left(-\textbf{k},-\omega\right).
\end{equation}

\noindent Thus all anisotropy does is to modify the coupling
constants but the local quantum criticality remains and the form of
the correlations at the critical point are still local in space and
power law in time. 

\section{Origin of Dissipation}

In addition to the bosonic modes associated with the order parameter, the low energy sector of the cuprates has fermionic degrees of freedom associated with electron near the fermi surface.
Quantum fluctuations of the order parameter correspond to flipping between states of the Ashkin-Teller model. From the mapping above of the Ashkin Teller model to an xy model with four fold anisotropy,
we note that a flip operator is  the quenched angular momentum operator. For the cuprates, the analogous operator can be shown to correspond to the curl of the fermionic current \cite{SHEK}.
The coupling between the fermionic degrees of freedom and the bosonic critical modes is of the form

\begin{equation}
H_{coup} \sim U(\textbf{r})\left(\nabla \times \textbf{j}\right)
\end{equation}

\noindent where $U$ is an operator which flips the state of the Ashkin Teller model (analogous to $\partial/\partial\theta$ in the continuous limit) and $\textbf{j}$ is the fermionic current.
Integrating out fermions leads to an Ohmic dissipative term in the action for the order parameter. The reason is that the current current correlation is the conductivity which goes as
$|\omega|$. Unlike Josephson junction arrays, the origin of dissipation in cuprates is the presence of both fermionic and bosonic degrees of freedom in the low energy long-wavelength theory.

\section{Coupling to other dissipative baths}

The form of dissipation considered so far is physically motivated to
reproduce a term linear in the gradient of the phase variable in its
equation of motion. Furthermore it can also be understood as arising
from dissipation via resistors connecting the islands on which the
phase degrees of freedom live. In principle one can also study the
effect of other forms of dissipation that are allowed by symmetry.
In particular, we can introduce a term which leads to suppression of
density in each grain instead of the phase difference between
nearest neighbor grains as in eqn.\ref{diss}.

\begin{equation}\label{densdiss1}
S_{n} = \gamma_{n}\sum_{i}\int d\tau d\tau
'{\left(n_{i}\left(\tau\right)-n_{i}\left(\tau
'\right)\right)^{2}\over {\left|\tau-\tau '\right|^{2}}}.
\end{equation}

To understand the effect of such a term on the dissipation driven critical point we consider the effect of integrating out frequency shells as before. We find that the coupling constant
$\gamma_{n}$ scales to zero since it has a dimension of $-2$. In other words such a form of dissipation is irrelevant near the local quantum critical point, while for a $z=1$ critical point such
a term is marginal.

A dissipative mechanism where the external bath directly couples to
the phase variable leads to an action of the form

\begin{equation}\label{densdiss2}
S_{\theta} = \gamma_{\theta}\sum_{i}\int d\tau d\tau
'{\left(\theta_{i}\left(\tau\right)-\theta_{i}\left(\tau
'\right)\right)^{2}\over {\left|\tau-\tau '\right|^{2}}}.
\end{equation}

\noindent The coupling $\gamma_{\theta}$ is marginal at the local
quantum critical point. A study of the critical properties showed
the existence of a line of fixed points controlling the transition
from the superconducting to the insulating state \cite{WAG1}, but
the approximation made were called into question \cite{AMS}. A weak coupling study of the model has been reported \cite{Tew_Ton} and a strong
coupling analysis will be the subject of future investigation

\section{Relationship to other models}

With the identification of the correct degrees of freedom we have been able to access the dissipation driven critical point that has been previously shown to exist within a perturbative
analysis. Furthermore, fourfold anisotropy does not alter the local criticality obtained at this quantum critical point. Thus the dissipative $xy$ model defines a new universality class. Classical
models which undergo a Gaussian phase transition from an ordered to a disordered phase will belong to this universality class provided their quantum generalization include Ohmic dissipation. In
this context we can look at the physics of six and eight vertex models.

The phase diagram of the six vertex model has two ordered phases, the orbital ferromagnet and the orbital antiferromagnet, and a power law phase. The high temperature phase is the power law
phase and at low temperatures undergoes a transition to one of the two ordered phases. It is known that there exists an essential singularity at the phase transition and the physics is similar
to the $xy$ model except that is inverted. The high temperature phase is power law correlated while the low temperature phase has finite correlations \cite{lieb}. Unlike the dissipative XY model
considered here, the inverted character of the phase transition, while allowing for a phase transition without a singularity in specific heat \cite{sudipsh}, is unlikely to belong to the
universality class of local criticality. However, the low temperature physics of this model including quantum generalization allowing for sources and sink is yet to be analyzed. The
classical eight vertex model on the other hand has in its parameter space a high temperature disordered phase which can undergo a KT transition to an ordered phase \cite{BAX}. Such models also
share the feature that the specific heat is nonsingular at the phase transition. Quantum generalization of this model with dissipation have not been studied. In particular it is not clear which
operators in the classical theory are equivalent to the warps defined in the continuum. A detailed analysis of the conditions for classical eight vertex models to exhibit local criticality is
beyond the scope of this paper.

Local quantum criticality closely related to the form proposed for the Cuprates \cite{mfl} and derived here has also been used to fit the spectral function near the quantum-critical point of a heavy-fermion \cite{SCHRODER} measured by neutron scattering. Some theoretical calculations \cite{SI} emphasizing the deconstruction of the single impurity Kondo effect near the quantum critical point \cite{MAEBASHI} have been performed. A general point based on the present work appears worth making: Local quantum criticality requires showing that near the quantum-critical point, there must be variables in terms of which the action for low energies is separable exactly into orthogonal parts which interact only in space without retardation in time, and those which are power law in time but interact only locally in space. It appears to us that only dissipation driven quantum critical points can have this property. At this point, it does not appear clear to us in terms of which variables, the general Action of the heavy fermion problem with dissipation, local Kondo interactions, interaction between local moments, magnetic order with attendant deconstruction of the Kondo effect, etc., may be cast into the simple structure due to which we have been able to solve the problems addressed in this paper.

\section{Summary}

In this paper we have provided a detailed analysis of the dissipation driven phase transition in the quantum 2d xy model. There exists a critical point in the phase diagram where the correlation
are shown to be local in space and power law in time. This is a new paradigm in critical phenomena which does not appear in the universality classes of classical dynamical critical phenomena \cite{HOHENBERG-HALPERIN}. The quantum disordering due to the proliferation of a new of class of topological defects, the warps, which
interact logarithmically in time but local in space allows such criticality. We have discussed that the results here are only asymptotically true and only valid for certain forms of dissipation. A verification of the results by quantum monte-carlo calculations is highly desirable.

A principal outcome of the paper is to derive the phenomenological assumptions of the marginal fermi-liquid theory with which the critical properties of the Cuprates in the anomalous metallic regime have been understood.

%\bibliography{at}% Produces the bibliography via BibTeX

\begin{thebibliography}{10}

\bibitem{aji1}
V.~Aji and C.M. Varma.
\newblock {\em Phys. Rev. Lett.}, 99:067003, 2007.

\bibitem{SC}
S.~Chakravarty, G.~Ingold, S.~Kivelson, and A.~Luther.
\newblock {\em Phys. Rev. Lett}, 56:2303, 1986.

\bibitem{MPAF}
M.P.A. Fisher.
\newblock {\em Phys. Rev. Lett}, 57:885, 1986.

\bibitem{ORR}
B.G. Orr, H.M. Jaeger, A.M. Golgman, and C.G. Kuper.
\newblock {\em Phys. Rev. Lett}, 56:378, 1986.

\bibitem{CL}
A.O. Caldeira and A.J. Leggett.
\newblock {\em Ann. Phys. (NY)}, 149:374, 1984.

\bibitem{SC1}
S.~Chakravarty, G.~Ingold, S.~Kivelson, and G.~Zimanyi.
\newblock {\em Phys. Rev. B}, 37:3283, 1988.

\bibitem{MPAF1}
M.P.A. Fisher.
\newblock {\em Phys. Rev. B}, 36:1917, 1987.

\bibitem{WAG}
K.~Wagenblast, A.~Otterlo, G.~Schon, and G.~Zimanyi.
\newblock {\em Phys. Rev. Lett.}, 79:2730, 1997.

\bibitem{Tewari}
S.Tewari, J.~Toner, and S.~Chakravarty.
\newblock {\em Phys. Rev. B}, 72:060505, 2005.

\bibitem{PAL}
P.~Goswami and S.~Chakravarty.
\newblock {\em Phys. Rev. B}, 73:094516, 2006.

\bibitem{GIL}
G.~Refael, E.~Demler, Y.~Oreg, and D.S. Fisher.
\newblock {\em Phys. Rev. B}, 75:014522, 2007.

\bibitem{FAQ}
B.~Fauque et~al.
\newblock {\em Phys. Rev. Lett.}, 96:197001, 2006.

\bibitem{mook}
H.~Mook.
\newblock In {\em Bulletin of American Physical Society March Meeting}, 2008.

\bibitem{greven}
Yuan Li, Victor Baledent, Neven Barisic, Philippe Bourges, Yongchan Cho, Benoit
  Fauque, Yvan Sidis, Guichuan Yu, Xudong Zhao, and Martin Greven.
\newblock In {\em Bulletin of American Physical Society March Meeting}, 2008.

\bibitem{bourges_pc}
P.~Bourges.
\newblock {\em Private Communication, Dec. 2008}.

\bibitem{AK}
A.~Kaminski et~al.
\newblock {\em Nature}, 416:610, 2002.

\bibitem{CMV}
C.M. Varma.
\newblock {\em Phys. Rev. B}, 55:14554, 1997.

\bibitem{CMV2}
C.M. Varma.
\newblock {\em Phys. Rev. B}, 73:155113, 2006.

\bibitem{simon}
M.E. Simon and C.M. Varma.
\newblock {\em Phys. Rev. Lett}, 49:1545, 1982.

\bibitem{BAX}
R.J. Baxter.
\newblock {\em Exactly Solved Models in Statistical Mechanics}.
\newblock Academic Press, 1982.

\bibitem{GRONSLETH}
M.S. Gronselth, T.B. Nilssen, E.K. Dahl, C.M. Varma, and A.~Sudbo.
\newblock {\em arXiv.org/cond-mat/0806.2665}.

\bibitem{AT}
J.~Ashkin and E.~Teller.
\newblock {\em Phys. Rev}, 64:178, 1943.

\bibitem{JOS}
J.V. Jose, L.P. Kadanoff, S.~Kirkpatrick, and D.R. Nelson.
\newblock {\em Phys. Rev. B}, 16:1217, 1977.

\bibitem{mfl}
C.M. Varma, P.B. Littlewood, S.~Schmitt-Rink, E.~Abrahams, and A.E.
  Ruckenstein.
\newblock {\em Phys. Rev. Lett.}, 63:1996, 1989.

\bibitem{SHEK}
Vivek Aji, Arkady Shekhter, and Chandra Varma.
\newblock {\em arXiv.org/cond-mat/0807.3741}.

\bibitem{SCHRODER}
A.~Schroder, G.~Aeppli, E.~Bucher, R.~Ramazashvili, and P.~Coleman.
\newblock {\em Phys. Rev. Lett.}, 80:5623, 1998.

\bibitem{SI}
Q.~Si, S.~Rabello, K.~Ingersent, and J.~Lleweilun Smith.
\newblock {\em Nature}, 416:610, 2002.

\bibitem{MW}
N.D. Mermin and H.~Wagner.
\newblock {\em Phys. Rev. Lett.}, 17:1113, 1966.

\bibitem{PCH}
P.C. Hohenberg.
\newblock {\em Phys. Rev.}, 158:383, 1967.

\bibitem{BER}
V.L. Berezinskii.
\newblock {\em Zh. Eksp. Teor. Fiz.}, 59:907, 1970.

\bibitem{KT}
J.M. Kosterlitz and D.J. Thouless.
\newblock {\em Jour. Phys. C}, 6:1181, 1973.

\bibitem{DH}
C.~Dasgupta and B.I. Halperin.
\newblock {\em Phys. Rev. Lett.}, 47:1556, 1981.

\bibitem{POL}
A.M. Polyakov.
\newblock {\em Nucl. Phys. B}, 120:429, 1977.

\bibitem{LEEFISHER}
D.H. Lee and M.P.A. Fisher.
\newblock {\em Phys. Rev. B}, 39:2756, 1989.

\bibitem{NAG}
N.~Nagaosa.
\newblock {\em Quantum Field Theory in Condensed Matter Physics, Sec. 5.2}.
\newblock Springer, 1999.

\bibitem{AHY}
P.W. Anderson, G.Yuval, and D.R. Hamann.
\newblock {\em Phys. Rev. B}, 1:4464, 1970.

\bibitem{BM}
A.J. Bray and M.A. Moore.
\newblock {\em Phys. Rev. Lett}, 89:247003, 2002.

\bibitem{SC3}
S.~Chakravarty.
\newblock {\em Phys. Rev. Lett}, 49:681, 1982.

\bibitem{footnote}
{\em It should be noted that even if the {\it bare} fugacity of the vortices
  and the warps may be given in terms of the parameters of the model by the
  same quantities, the flow equations of the two are quite different, so that
  vortices and warps remain independent variables.}

\bibitem{LEG}
A.J. Leggett, S.~Chakravarty, A.T. Dorsey, M.P.A. Fisher, A.~Garg, and
  W.~Zwerger.
\newblock {\em Rev. Mod. Phys}, 59:1, 1987.

\bibitem{Kosterlitz}
J.M. Kosterlitz.
\newblock {\em J. Phys. C}, 7:1046, 1974.

\bibitem{GAS}
A.~Vishwanath P.~Ghaemi and T.Senthil.
\newblock {\em Phys. Rev. B}, 72:024420, 2005.

\bibitem{IZ}
J.B. Zuber and C.~Itzykson.
\newblock {\em Phys. Rev. D}, 15:2875, 1977.

\bibitem{KD}
L.P. Kadanoff and A.C. Brown.
\newblock {\em Ann. of Phys.}, 121:318, 1979.

\bibitem{WAG1}
K.~Wagenblast, A.~Otterlo, G.~Schon, and G.~Zimanyi.
\newblock {\em Phys. Rev. Lett.}, 78:1779, 1997.

\bibitem{AMS}
A.~Vishwanath, J.E. Moore, and T.~Senthil.
\newblock {\em Phys. Rev. B}, 69:054507, 2004.

\bibitem{lieb}
E.~Lieb.
\newblock {\em Phys. Rev. Lett.}, 18:1046, 1967.

\bibitem{sudipsh}
S.~Chakravarty.
\newblock {\em Phys. Rev. B}, 66:224505, 2002.

\bibitem{MAEBASHI}
H.~Maebashi, K.~Miyake, and C.M. Varma.
\newblock {\em Phys. Rev. Lett.}, 95:207207, 2005.

\bibitem{HOHENBERG-HALPERIN}
P.C. Hohenberg and B.I. Halperin.
\newblock {\em REv. Mod. Phys.}, 49:435, 1977.

\bibitem{Tew_Ton}
S. Teawri and J. Toner.
\newblock {\em Europhys. Lett.}, 74:341, 2006.

\end{thebibliography}
%\bibliographystyle{unsrt}

\end{document}